\documentclass[journal]{IEEEtran}

\usepackage[dvipsnames]{xcolor}
\usepackage{booktabs}
\usepackage{float}
\usepackage{array}                
\usepackage{amsmath, amssymb, bm}  
\usepackage{graphicx}    
\usepackage{tabularx}
\usepackage[table]{xcolor}
\usepackage{multirow}
\usepackage{cite}
\usepackage{soul}

\ifCLASSINFOpdf

\else

\fi

\begin{document}

\title{Transceiver-Integrated BD-RIS: Wave-Domain Signal Processing for Sustainable and Inclusive 6G}

\author{Mahmoud~Raeisi,~\IEEEmembership{Member,~IEEE,}
        Ayoub~Ammar~Boudjelal,~\IEEEmembership{Student~Member,~IEEE,}
        Henk~Wymeersch,~\IEEEmembership{Fellow,~IEEE,}
        Ertugrul~Basar,~\IEEEmembership{Fellow,~IEEE,}
        and~Huseyin~Arslan,~\IEEEmembership{Fellow,~IEEE}
\thanks{Mahmoud Raeisi, Ayoub~Ammar~Boudjelal, and Huseyin~Arslan are with the Department of Electrical and Electronics Engineering, Istanbul Medipol University, 34810 Istanbul, Turkey. (e-mail: mahmoud.raeisi@medipol.edu.tr; ayoub.ammar@std.medipol.edu.tr; huseyinarslan@medipol.edu.tr)}
\thanks{Henk Wymeersch is with the Department of Electrical Engineering, Chalmers University of Technology, 41296 Gothenburg, Sweden (e-mail: henkw@chalmers.se).}
\thanks{Ertugrul Basar is with the Tampere Wireless Research Center, Department of Electrical Engineering, Tampere University, 33720 Tampere, Finland, on leave from the Department of Electrical and Electronics Engineering, Koc University, 34450 Sariyer, Istanbul, Turkey (email: ertugrul.basar@tuni.fi).}
\thanks{This work was supported in part by the EU Project RAI-6GREEN under the EUREKA CELTIC-NEXT Project, C2023/1-9, funded by TÜBİTAK under the TEYDEB 1509 Program, Project No. 9230046, and in part by the Swedish Research Council (VR grant 2022-03007).}
}

\maketitle

\begin{abstract}
The shift toward sixth-generation (6G) wireless communications demands transceiver architectures that simultaneously support high-data-rate communications, pervasive sensing, and sub-meter-level localization. Beyond these performance targets, 6G systems are also expected to align with long-term societal goals, including sustainability and inclusiveness. Conventional radio designs, however, remain heavily reliant on digital baseband processing, whose cost, power consumption, and computational complexity scale unfavorably with increasing array size and carrier frequency, making them poorly aligned with these emerging requirements. Beyond-diagonal reconfigurable intelligent surfaces (BD-RISs) introduce a new paradigm by enabling direct manipulation of electromagnetic waves in the analog domain. This article presents BD-RIS as a wave-domain analog processing unit embedded within the transceiver aperture. By migrating linear signal processing functions from the digital baseband to the wave domain, BD-RISs significantly reduce computational load and energy consumption, enabling scalable and sustainable operation for extra-large antenna array systems. Owing to their ability to jointly provide high operational flexibility, modularity, and energy-efficient analog processing, transceiver-integrated BD-RISs offer a compelling architectural trade-off and emerge as a strong candidate for next-generation wireless transceivers.
\end{abstract}

\IEEEpeerreviewmaketitle


\section{Introduction: A Paradigm Shift Toward Analog Signal Processing} 
\label{Sec: Introduction}

\begin{figure*}
    \centering
    \includegraphics[width=\textwidth]{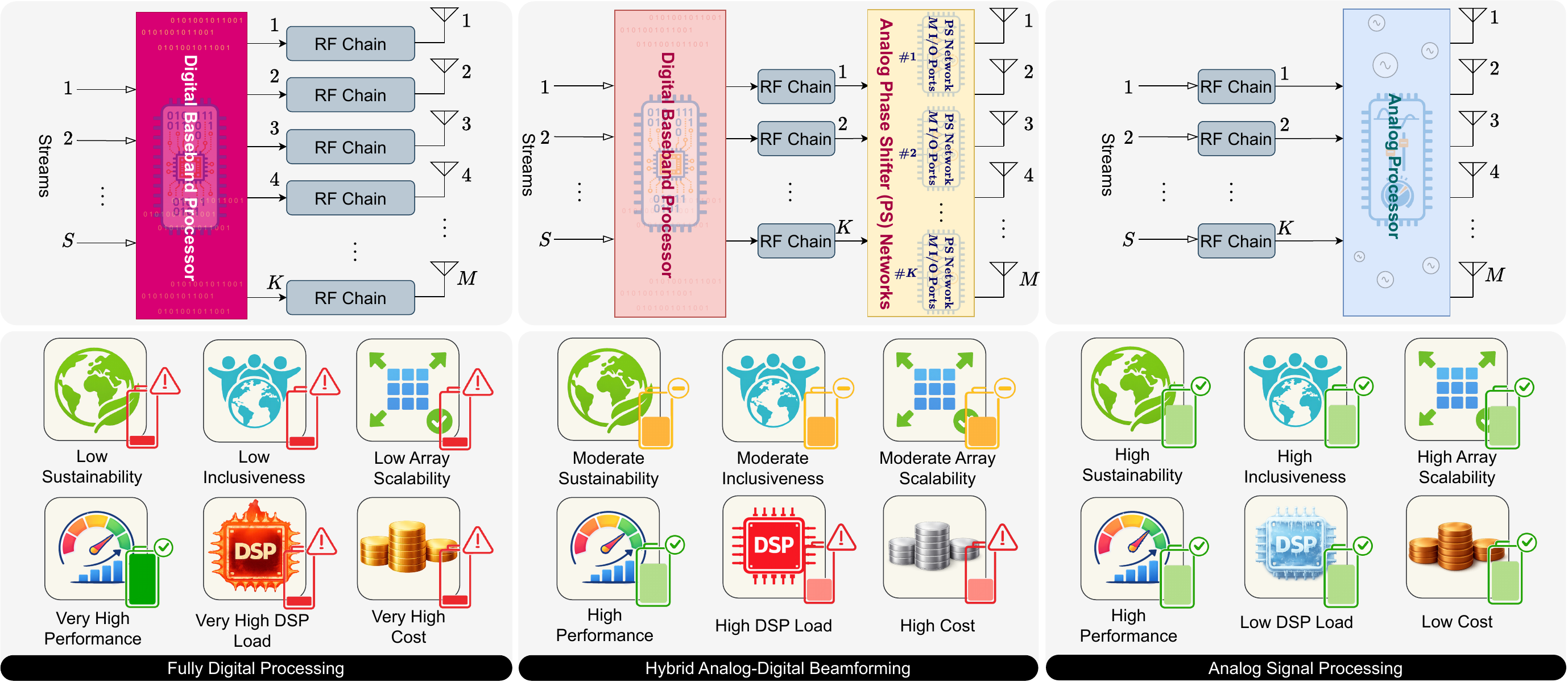}
    \caption{Architectural partitioning of signal processing functionality in multi-antenna transceivers: (left) fully digital architectures, (center) hybrid analog–digital beamforming architectures, and (right) analog processing architectures. The qualitative indicators highlight the favorable trade-offs offered by analog processing architectures, positioning them as a strong class of transceivers for 6G and beyond.}
    \label{fig:Array_Architecture}
\end{figure*}

The move toward sixth-generation (6G) wireless networks brings unprecedented demands on wireless systems, including both communication and sensing applications. These demands must be met under the stringent key value indicators (KVIs) of sustainability, inclusiveness, and trustworthiness defined for 6G systems \cite{10872853}. Fully digital multiple-input multiple-output (MIMO) architectures offer maximum performance and flexibility by carrying out signal processing entirely in the digital baseband, with independent control of the signal at each antenna element \cite{11281878}. Nonetheless, such architectures require a dedicated radio-frequency (RF) chain and a high-resolution analog-to-digital/digital-to-analog converter (ADC/DAC) for each antenna element. As arrays scale toward extra-large MIMO with thousands of antenna elements, the resulting linear growth in RF chains leads to prohibitive cost and power consumption, rendering fully digital designs increasingly impractical. The structural features of fully digital architectures, alongside their qualitative characteristics, are summarized in the left panel of Fig. \ref{fig:Array_Architecture}.

To alleviate these limitations, the hybrid analog–digital (A/D) beamforming architecture is introduced, shifting spatial beamforming into the RF domain through active analog phase-shifter networks \cite{11301914}. Hybrid beamforming maintains much of the performance of fully digital designs while significantly reducing the number of RF chains. However, the phase-shifter network itself becomes increasingly bulky and power-hungry as array sizes and the number of RF chains grow \cite{11301914}. As a result, hybrid beamforming architecture struggles to meet the sustainability and inclusiveness requirements envisioned for future 6G transceivers, motivating the search for fundamentally new transceiver architectures. The architectural limitation in scaling phase shifter networks, along with its effects on various metrics and values, is illustrated in the center panel of Fig.~\ref{fig:Array_Architecture}.

Given that scalability is inherent to both fully digital and hybrid A/D architectures, a paradigm shift in MIMO transceiver design is necessary to address this limitation. Recent works have proposed shifting signal processing into the analog domain, going beyond conventional hybrid A/D designs, toward architectures capable of performing more general analog transformations \cite{11281878,11301914,raeisi2024efficient,10515204}. This structural shift reduces arithmetic load in the baseband domain and relaxes ADC/DAC resolution requirements\textcolor{black}{, while introducing new configuration complexity in the underlying analog processing network}. As a result, such analog processing designs can scale more gracefully to ultra-massive arrays. The right panel of Fig. \ref{fig:Array_Architecture} visually illustrates the architectural features of analog processing transceivers and their qualitative enhancements. A side-by-side comparison among the three panels of Fig. \ref{fig:Array_Architecture} reveals that analog signal processing transceivers offer a more balanced trade-off among performance, cost, scalability, and the 6G KVIs of sustainability and inclusiveness.

In view of the favorable trade-offs offered by analog-domain processing, several architectures have recently been proposed, including microwave linear analog computers (MiLACs), reconfigurable intelligent surface (RIS)-integrated arrays, stacked intelligent metasurfaces (SIMs), and transceiver-integrated beyond-diagonal RIS (BD-RIS) architectures. MiLAC performs analog processing through wired microwave networks embedded within the transceiver front~end, aiming to reduce digital complexity while preserving high performance \cite{11281878}. In a complementary direction, electromagnetic (EM) wave manipulation has been investigated as a means to realize analog signal processing at the transceivers. RISs, introduced initially for environment-level control of wave propagation, have also been considered for integration at the transceiver aperture as passive beamforming structures \cite{11281878}. 
\textcolor{black}{Along this line, multi-layer RIS-based receiver concepts have been introduced to improve integrated communication and energy transfer capabilities \cite{10520169}.}
To enable more advanced wave-domain processing within the transceiver, SIMs have been introduced that exploit cascaded EM interactions across multiple RIS layers \cite{10515204}. While MiLAC offers the highest degree of freedom in realizing linear signal processing functions, SIM provides a modular add-on form factor. In contrast, the concept of transceiver-integrated BD-RIS offers a middle ground between these two directions by enabling analog signal processing through interconnected surface elements while retaining a modular structure that can be attached to existing base station (BS) hardware. As a result, it represents an attractive architecture from both technological and economic perspectives.

This article presents transceiver-integrated BD-RIS as a new transceiver architecture that enables linear signal processing directly in the analog wave domain. This paradigm shift paves the way for scalable, sustainable, and inclusive transceivers aligned with 6G societal goals. Despite the high potential of BD-RISs to act as analog processors co-located with transceivers, existing magazine and survey articles portray BD-RISs mainly as environment-mounted platforms for shaping the wireless propagation environment \cite{li2023reconfigurable,11311540}. In contrast, this article is the first to discuss the BD-RIS architecture as an analog processing unit that can be implemented at the transceiver aperture, highlighting its potential for analog signal processing and presenting comparative insights and trade-offs with other analog processing architectures.

\begin{figure*}
    \centering
    \includegraphics[width = \textwidth]{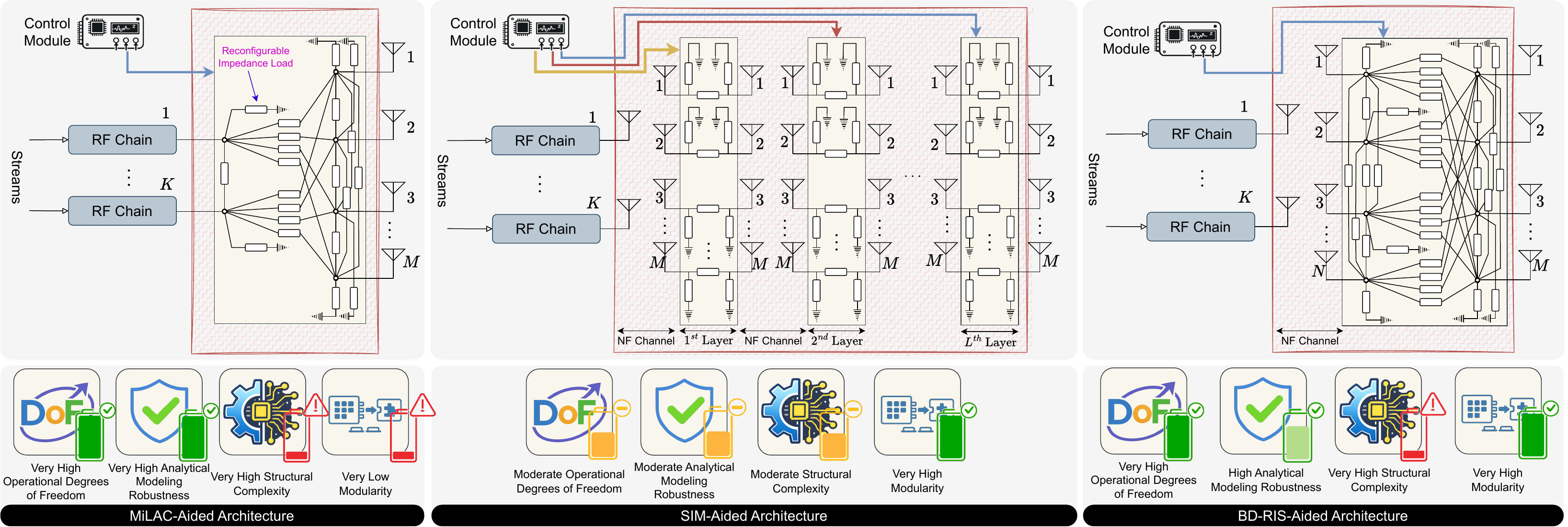}
    \caption{Comparative illustration of MiLAC-aided, SIM-aided, and BD-RIS-aided analog processors. The three architectures differ in how analog signal transformations are implemented: (left) circuit-domain processing via MiLAC networks, (center) wave-domain processing through cascaded SIM layers, and (right) wave-domain processing enabled by BD-RISs with interconnected impedance networks. The qualitative indicators highlight the associated trends in operational degrees of freedom, analytical modeling robustness, structural complexity, and modularity.}
    \label{fig:Analog_Front_End}
\end{figure*}

\section{Analog Signal Processing Architectures for 6G Transceivers}\label{Sec: Front-End-Centric Transcievers} 

This section surveys the main analog signal processing architectures proposed for 6G transceivers, highlighting their operating principles, capabilities, and key implementation trade-offs. We organize these architectures into two broad categories based on their underlying operating principles. MiLACs operate within the transceiver circuit chain, between the RF chain and the antenna array, where analog processing is carried out by manipulating electrical signals. The second category relies on metasurfaces, which manipulate the EM wave directly at the antenna aperture.

\subsection{Circuit-Domain Analog Signal Processing}

MiLACs represent a recently proposed approach for analog signal processing that operates in the microwave circuit domain. They are implemented as multiport reconfigurable microwave networks composed of tunable impedance components that process RF signals as they propagate through the network (Fig. \ref{fig:Analog_Front_End}, left panel) \cite{11281878}. Unlike digital baseband processors that manipulate quantized samples, MiLACs apply the desired linear transformations directly in the analog domain. By connecting a limited number of RF chains to an extra-large set of antenna ports, MiLACs can implement linear precoding or combining operations in the analog domain, thereby alleviating the burden on digital processing. Existing studies indicate that fully-connected MiLAC architectures can achieve a high performance, close to fully digital systems, positioning them as a viable realization path for front~end analog processing \cite{11281878}.


\subsection{Wave-Domain Analog Signal Processing}

Wave-domain processing architectures manipulate EM waves at the antenna aperture, in the immediate near field region of the transceiver. Two representative realizations of wave-domain processing architectures are SIMs and transmissive BD-RISs, co-located in the transceivers, as illustrated in the center and right panels of Fig. \ref{fig:Analog_Front_End}, respectively. In both cases, appropriately configured impedance loads shape the incident wavefront, enabling wave-domain signal processing \cite{11301914, 10515204}.

SIMs achieve analog processing through multiple transmissive RIS layers arranged in cascade, as illustrated in the center panel of Fig. \ref{fig:Analog_Front_End} \cite{10515204}. The successive interactions across layers implement a sequence of controllable transformations on the propagating wave, enabling advanced wave-domain signal processing operations beyond simple beam steering \cite{10515204}. 
\textcolor{black}{Photographs of two SIM prototypes are presented in \cite[Fig.~1]{10557708}.}
BD-RISs, on the other hand, generalize conventional diagonal RIS architectures by equipping the surface with a reconfigurable impedance network that interconnects all elements \cite{li2023reconfigurable}. Rather than each element responding independently to the incident wave, the impedance network supports distributed EM responses governed by the configuration of this underlying impedance network. Hence, the surface can realize wave-domain signal processing with higher degrees of freedom and less power loss due to reduced multiplicative path loss compared to SIM \cite{11301914, raeisi2024efficient}. An illustration of transceiver-integrated BD-RIS is shown in the right panel of Fig. \ref{fig:Analog_Front_End}. \textcolor{black}{Here, the end-to-end channel is modeled as the cascade of the transceiver–BD-RIS and BD-RIS–user propagation links, which preserves reciprocity. A photograph of a BD-RIS prototype is presented in \cite[Fig.~14]{11195964}.}

\subsection{Comparative Perspective and Architectural Insights} \label{Sec: Comparitive Perspective}

From a functional standpoint, MiLACs, SIMs, and BD-RISs all enable analog signal processing, but they do so with different levels of reconfigurability, modeling fidelity, implementation complexity, and modularity. These differences, together with the structural features of each architecture, are illustrated in Fig. \ref{fig:Analog_Front_End}.

\begin{enumerate}
    \item \textit{Degrees of freedom in signal processing:} MiLAC and BD-RIS architectures provide a high degree of freedom for implementing analog signal processing tasks. This capability stems from their fully adjustable reconfigurable impedance network, enabling the realization of a broad class of linear transformations in the analog domain. SIM architectures also operate in the wave domain through multiple stacked metasurface layers; however, only the metasurface element responses are typically adjustable, whereas the inter-layer transmission coefficients are not tunable \cite{10515204}. In practice, these coefficient terms may also deviate from analytical models because of hardware losses, parasitics, and fabrication tolerances \cite{10515204}. Collectively, these structural and implementation constraints restrict the flexibility of realizable operations and, hence, reduce the effective degrees of freedom of SIMs for a target analog signal processing operation.

    \item \textit{Modeling accuracy and robustness:} Both SIM and BD-RIS architectures involve near field wave propagation between physical structures, which in practice requires channel calibration to ensure accurate modeling. In SIM, \textcolor{black}{as described in \cite[Eq.~(14)]{10557708},} the presence of multiple cascaded layers gives rise to several near field channels; therefore, imperfections across layers accumulate, and modeling errors compound, making calibration more demanding \cite{10515204}. In contrast, BD-RIS involves only a single near field channel between the antenna array and the BD-RIS aperture \textcolor{black}{\cite[Eq.~(8)]{11301914}}, which reduces the extent of error accumulation and simplifies the associated calibration task. On the other hand, MiLAC \textcolor{black}{is embedded within the transceiver circuitry and can be modeled as a reconfigurable microwave network \cite[Eq.~(1)]{11281878}. Since it} does not introduce an intermediate near field propagation channel between separate structures, it inherently avoids the need for such calibration.

    \item \textit{Structural complexity:} SIMs rely on a conceptually simple structure but incur optimization-driven complexity. Stacking multiple layers introduces strong inter-layer coupling, so achieving good performance requires joint multi-layer optimization together with additional layer-to-layer calibration. Fully-connected BD-RIS and MiLAC, in contrast, entail higher circuit complexity: their reconfigurable impedance networks require many tunable loads and sophisticated control circuitry, leading to a high-dimensional configuration space and computationally demanding optimization. As a result, fully flexible implementations of these architectures may become impractical. Structured realizations, such as group-connected or partially connected layouts, sparse impedance networks, or partially fixed loads with limited reconfiguration, can instead be adopted \textcolor{black}{to reduce configuration overhead and avoid merely shifting computation from real-time DSP to analog configuration, while retaining performance close to that of fully-connected networks.}

    \item \textit{Modularity and ease of integration:} SIM and BD-RIS architectures offer a key practical advantage in modularity and ease of deployment, as they can be introduced as add-on modules to existing transceivers without modifying the core RF hardware. This makes them well-suited for retrofitting legacy infrastructure, incremental upgrades, and \textcolor{black}{add-on integration}, with reduced upgrading cost. On the other hand, MiLAC is tightly embedded within the transceiver hardware through electrical connectors and typically requires co-design with the RF front~end. Therefore, different BS configurations require different MiLAC hardware designs. Such vendor-specific integration increases manufacturing and inventory complexity, complicates interoperability and future upgrades, and ultimately limits its cost efficiency and practicality for large-scale deployment.
\end{enumerate}

The features of each architecture are visually summarized in Fig. \ref{fig:Analog_Front_End}, illustrating that transceiver-integrated BD-RIS architectures offer a balanced trade-off across the discussed dimensions. Accordingly, the remainder of this article focuses on transceiver-integrated BD-RIS as a promising architecture for next-generation wireless communication systems.

\section{Transceiver-Integrated BD-RIS: Toward 6G Societal Goals}

\begin{figure*}
    \centering
    \includegraphics[width=\textwidth]{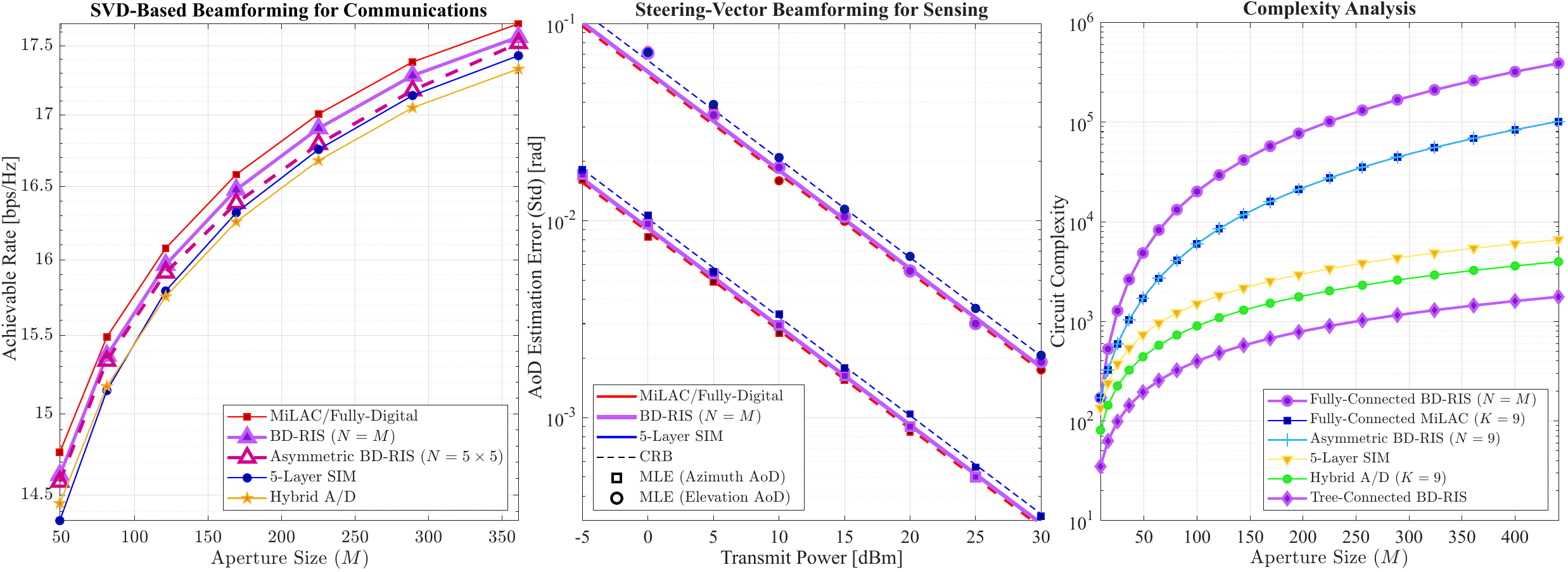}
    \caption{Performance comparison of transceiver architectures, illustrating communication performance (left), sensing accuracy (center), and the scaling of circuit complexity with the number of elements (right). All scenarios are evaluated in the downlink, where the BS employs a large antenna array and the UE is equipped with a single antenna. The UE observes an effective aperture of $M = 9 \times 9$ elements. For the sensing scenario, both the theoretical Cramér--Rao bound (CRB) and the maximum likelihood estimator (MLE) are used to evaluate the angle-of-departure (AoD) estimation accuracy.}
    \label{fig:Sims}
\end{figure*}

This section presents a quantitative insight into transceiver-integrated BD-RIS in communication and sensing applications, highlighting how its architectural design shapes trade-offs between performance and complexity compared to the benchmark architectures.

{\color{black}
\subsection{Representative Case Study for 6G Transceiver Evolution}

To illustrate the role of transceiver-integrated BD-RIS in 6G transceiver design, we consider a representative point-to-point downlink scenario in which a BS with a large antenna array serves a single-antenna UE observing an effective aperture of size $M$ elements. The communication and sensing scenarios follow the parameter settings in \cite{11301914} and \cite{raeisi2024efficient}, respectively. 
In this setting, MiLAC achieves performance equivalent to the fully digital architecture \cite{11281878}, and thus provides a reference benchmark for assessing other analog architectures while highlighting the potential of analog processing to alleviate the energy and scalability limitations of fully digital transceivers. 
The communication, sensing, and circuit complexity aspects are detailed in the following subsections, with the corresponding case study results illustrated in Fig.~\ref{fig:Sims}. To complement these quantitative insights, Table \ref{Tab: Comparison} provides a qualitative comparison of the considered architectures across deployment vision, scalability, processing, and cost dimensions, offering a holistic view of their respective trade-offs.
} 

\subsection{Wave-Domain Precoding for Communication}
\label{Sec:CaseStudy_Comm}

In the communication scenario, beamforming is based on the dominant right singular vector of the channel matrix, obtained via singular value decomposition (SVD). This precoding vector entails both amplitude and phase adjustment; MiLAC and BD-RIS can directly implement such a precoder through their interconnected impedance networks, whereas SIM must approximate it by solving an optimization problem to determine the phase shifts of its stacked layers \cite{10557708, li2025stacked}. 
\textcolor{black}{As shown in the left panel of Fig. \ref{fig:Sims}, MiLAC achieves the highest performance, serving as a benchmark for optimal precoding. However, this gain comes at the cost of poor modularity, which limits its practicality for large-scale deployment. In contrast, transceiver-integrated BD-RIS achieves a performance close to MiLAC while maintaining a modular add-on structure. This highlights its ability to strike a favorable balance between performance and implementation flexibility. The slight performance loss is mainly attributed to the multiplicative path loss introduced by the cascaded channel. SIM offers a level of modularity comparable to BD-RIS; however, its multi-layer structure and reliance on approximations of the precoding matrix result in a more noticeable performance degradation. Finally, hybrid A/D exhibits the lowest performance among the considered schemes, due to its unit-modulus constraint and the lack of amplitude control across the array.}

\subsection{Wave-Domain Probing for Sensing}
\label{Sec:CaseStudy_Sens}

In the sensing scenario, channel state information is not available, and a sweeping stage is required to probe the channel. A codebook of unit-modulus steering vectors is employed for environment sweeping, and sensing is then performed based on the received signal responses \cite{raeisi2024efficient}. As shown in the center panel of Fig. \ref{fig:Sims}, the sensing accuracy of BD-RIS closely follows that of MiLAC, indicating the high potential of transceiver-integrated BD-RIS in sensing scenarios. \textcolor{black}{In contrast, SIM exhibits slightly degraded performance, which can be attributed to its multi-layer structure that introduces more pronounced multiplicative path loss compared to the single-layer BD-RIS architecture.} Noting that owing to the unit-modulus nature of the sweeping codebook, MiLAC and hybrid A/D achieve essentially identical sensing performance; therefore, the hybrid A/D architecture is not considered separately.

\subsection{Complexity and Scalability}
\label{Sec:CaseStudy_Complx}

Besides performance and modularity, complexity is another important factor, as it affects the feasibility and scalability of each transceiver design. Circuit complexity is determined by the number of reconfigurable components, which also reflects the number of optimization variables and the associated control overhead. For fully-connected BD-RIS and MiLAC, an upper bound on the number of required reconfigurable components can be obtained straightforwardly through combinatorial counting.  
As depicted in Fig. \ref{fig:Analog_Front_End}, both MiLAC and BD-RIS follow the same underlying circuit design principle. However, MiLAC is physically connected to the RF chains through electrical connectors; that is, the number of ports in the RF-chain-facing sector is $K$, which is typically much smaller than the number of ports in the environment-facing sector $M$. In contrast, both sectors of BD-RIS are equipped with $M$ elements, which requires a larger number of reconfigurable impedance loads to interconnect all elements. Consequently, as shown in the right panel of Fig. \ref{fig:Sims}, a fully-connected BD-RIS exhibits the highest circuit complexity.

\begin{table*}[t]
\centering
\caption{Comparative overview of representative transceivers across deployment vision, scalability, modeling, and cost dimensions, illustrating the qualitative trade-offs among performance, modularity, structural complexity, and energy efficiency. \\
$M$: effective aperture size from the UE perspective; $N$: Number of BD-RIS elements facing the BS; $K$: Number of RF chains; $L$: Number of SIM layers.}
\renewcommand{\arraystretch}{1.25}

\begin{tabular}{|c||c||c|c|c|c|c|}
\hline
\textbf{Category} & \textbf{Attribute} 
& \textbf{Fully Digital} 
& \textbf{Hybrid A/D} 
& \textbf{SIM} 
& \textbf{MiLAC} 
& \textbf{BD-RIS} \\
\hline \hline

\multirow{3}{*}{\rotatebox{0}{\textbf{Deployment/Vision}}}
& \textit{Sustainability} & Low & Moderate & High & High & High \\ \cline{2-7}
& \textit{Inclusiveness} & Low & Moderate & High & Moderate & High \\ \cline{2-7}
& \textit{Modularity} & Low & Low & High & Low & High \\
\hline \hline

\multirow{2}{*}{\rotatebox{0}{\textbf{Scalability/Capability}}}
& \textit{Array Scalability} & Low & Moderate & High & High & High \\ \cline{2-7}
& \textit{Degrees of Freedom} & Very High & Moderate & High & Very High & Very High \\
\hline \hline

\multirow{5}{*}{\rotatebox{0}{\textbf{Processing/Modeling}}}
& \textit{Processing Domain} & Baseband & Baseband + RF & EM Wave & Analog Circuit & EM Wave \\ \cline{2-7}
& \textit{DSP Load} & Very High & High & Low & Low & Low \\ \cline{2-7}
& \textit{Circuit Complexity (Upper Bound)} & N/A & $\mathcal{O}(KM)$ & $\mathcal{O}(LM)$ & $\mathcal{O}\left((M+K)^2\right)$ & $\mathcal{O}((M+N)^2)$ \\ \cline{2-7}
& \textit{Circuit Design Flexibility} & Very Low & Moderate & Low & High & Very High \\ \cline{2-7}
& \textit{Robust Analytical Modeling} & Very High & Very High & Moderate & Very High & High \\
\hline \hline

\multirow{2}{*}{\rotatebox{0}{\textbf{Cost/Energy}}}
& \textit{Power Consumption} & Very High & Moderate & Low & Low & Low \\ \cline{2-7}
& \textit{Cost} & Very High & High & Low & Moderate & Low \\
\hline

\end{tabular}
\label{Tab: Comparison}
\end{table*}

Nonetheless, the strong modularity of BD-RIS makes its design largely independent of the number of RF chains, offering substantial flexibility across different deployment scenarios. In particular, for transceiver integration applications, the number of BD-RIS elements in the active-antenna-facing sector can be reduced independently of the environment-facing sector. This leads to an asymmetric arrangement of elements across the two sectors, which significantly lowers circuit complexity with only a modest performance penalty. The resulting performance–complexity trade-off of this upgraded BD-RIS, referred to as asymmetric BD-RIS, is illustrated in Fig. \ref{fig:Sims}. Here, $N$ denotes the number of elements in the active-antenna-facing sector. 
\textcolor{black}{For comparison, SIM and hybrid A/D exhibit moderate circuit complexity, as they rely on simpler structural designs without fully-connected impedance networks. However, this reduced complexity comes at the expense of limited flexibility compared to BD-RIS and MiLAC architectures.}
Designing low-complexity circuits for both MiLAC and BD-RIS remains an open research direction. \textcolor{black}{To explicitly demonstrate how structured low-complexity BD-RIS architectures can alleviate the configuration and scalability challenges of fully-connected designs, this article considers a representative graph-theoretic realization, namely tree-connected BD-RIS \cite{10857964}, which not only reduces circuit complexity but also enables fast reconfiguration of the underlying impedance network.} As shown in the right panel of Fig. \ref{fig:Sims}, tree-connected BD-RIS can even achieve lower complexity than hybrid A/D, offering a promising vision for future low-complexity transceiver-integrated BD-RIS designs.

\subsection{Lessons Learned}

\textcolor{black}{The results in Fig. \ref{fig:Sims} collectively highlight the fundamental trade-offs among communication performance, sensing accuracy, and circuit complexity. In particular, BD-RIS achieves a favorable balance, maintaining competitive performance while enabling scalable and modular implementations. Complementing these observations,} Table \ref{Tab: Comparison} summarizes and compares the main features of the considered transceiver architectures elaborated in this article. The table further reveals that transceiver-integrated BD-RIS strikes a balanced trade-off, incorporating the modularity and scalability of SIM with the robust modeling capability and high flexibility of MiLAC. 

\section{Wave-Domain Functions for 6G Wireless Networks}

\begin{figure}
    \centering
    \includegraphics[width=\columnwidth]{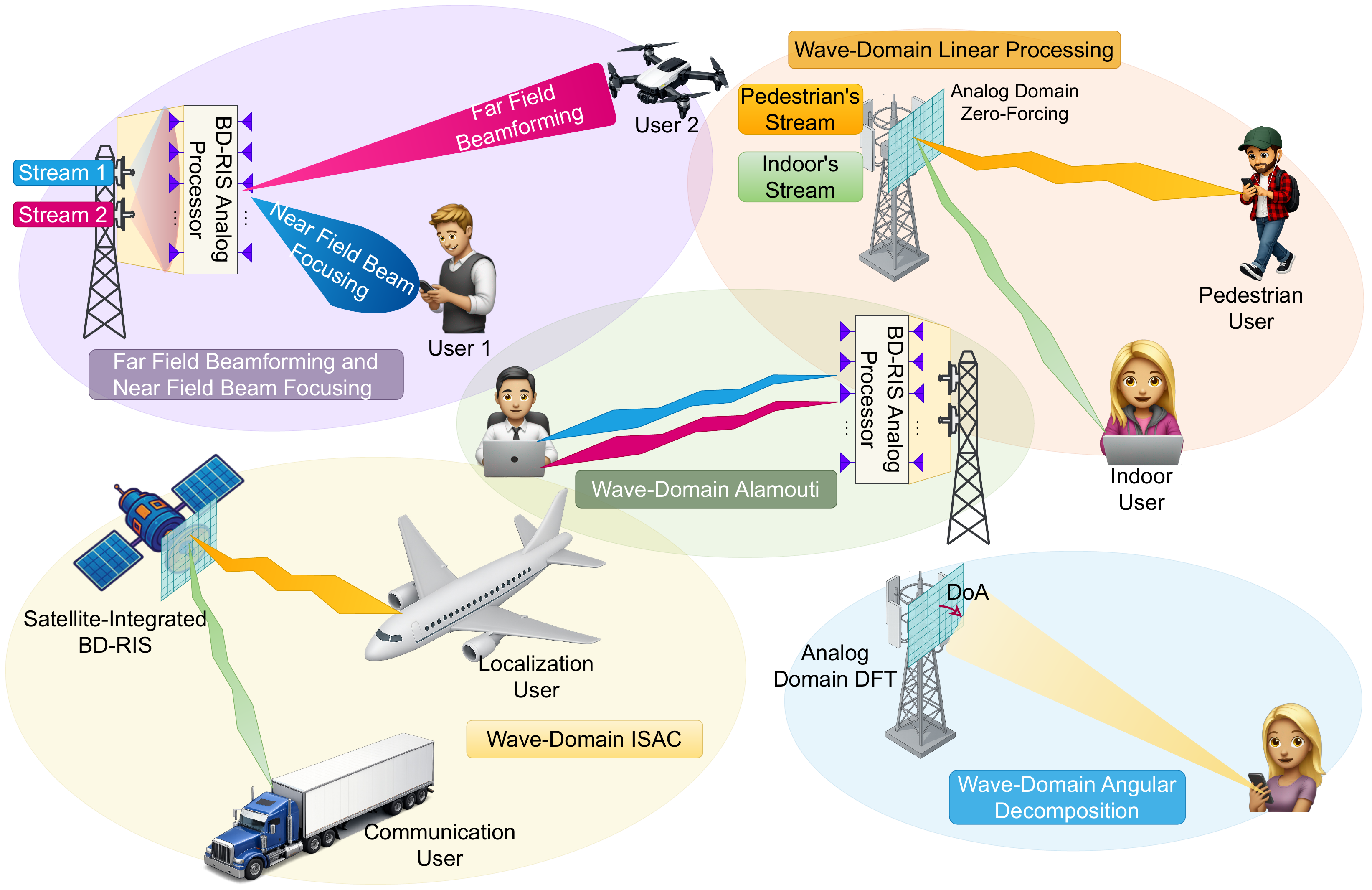}
    \caption{Illustration of representative wave-domain processing capabilities enabled by transceiver-integrated BD-RIS in 6G wireless networks, highlighting analog-domain manipulation of wavefronts to support beamforming, transform-domain processing, interference management, and joint communication, sensing, and localization.}
    \label{fig: General Scenario}
\end{figure}

This section presents a representative set of functions that can be migrated from the digital baseband to the analog wave domain, thereby offloading power-hungry arithmetic operations from the digital processor. These examples are not exhaustive; rather, they illustrate a flexible and extensible paradigm that can support a wide range of operations in both communication and sensing. An overview of representative wave-domain functions enabled by transceiver-integrated BD-RIS is illustrated in Fig. \ref{fig: General Scenario}.

\subsection{Beamforming and Beam Focusing} 

Leveraging an interconnected impedance network, BD-RIS enables highly flexible wave manipulation by controlling both the amplitude and phase of the incident signal across the array aperture. This high degree of freedom enables the transmitter to form the beam at different angles in the far field or focus it on a specific spot in the near field in the serving environment, without relying on traditional phase shifters in the RF or power-hungry digital processing in the baseband. Moreover, the circuit complexity of BD-RIS is independent of the number of RF chains. Therefore, unlike fully digital, hybrid beamforming, and MiLAC architectures, its complexity does not grow with the number of data streams, provided that the aperture is sufficiently large to support these independent streams.

\subsection{Wave-Domain Linear Processing}

In current wireless systems, matrix-based precoding and combining are implemented digitally using arithmetic operations. Nevertheless, popular functions such as SVD-based precoding and zero-forcing can be realized in the wave domain, avoiding power-hungry baseband arithmetic. In addition to reducing energy consumption and computational load, wave-domain operations avoid clocked digital processing, thereby eliminating baseband arithmetic latency and enabling extremely low processing delay that aligns well with 6G requirements. Besides, parallel analog computing can be enabled on BD-RIS by implementing an effective combined matrix of multiple functions, further highlighting the high capability of wave-domain processing.

\subsection{Integrated Sensing and Communications (ISAC)} 

In the next generation of wireless communication, sensing is considered to be natively integrated. However, embedding sensing functionality within communication systems further increases the computational burden. Here, transceiver-integrated BD-RIS can play a significant role by offloading part of this burden to the wave domain through analog processing. Specifically, simultaneous sensing and communication involves strong interference between these two functions. Such interference can be mitigated directly in the wave domain via analog processing functions implemented on the BD-RIS. In addition, BD-RIS can potentially host linear operators, such as discrete Fourier transform (DFT)-type impedance patterns, to perform part of the estimation and decomposition tasks in the analog domain, thereby reducing arithmetic and processing burdens at the digital baseband.

\subsection{Transform-Domain Channel Conditioning}

In wideband and multicarrier systems, a BD-RIS can manipulate the spatial and frequency structure of the signal directly in the EM domain \cite{10857964}. Rather than relying exclusively on digital baseband processing, the reconfigurable impedance network of the BD-RIS enables physically realized transform-domain operations that shape the incoming wave before sampling. 
Representative operations include the DFT, which provides angular decomposition of far field wavefront \cite{10557708}, and discrete Fresnel transforms that control near field curvature by imposing quadratic phase profiles across the aperture \cite{10217152}. These operations organize the incident wavefront into structured spatial and frequency components, yielding a more uniform and well-conditioned signal prior to sampling. This wave-domain structuring inherently acts as a form of analog channel conditioning: by shaping both amplitude and phase responses in the EM domain, the surface restructures the effective wideband channel across subbands, enhances dominant propagation components, and alters the impact of multipath reflections \cite{li2025stacked,10857964}. As a result, frequency-selective distortions and multipath-induced non-uniformities are mitigated prior to digital processing. By conditioning the channel in the wave domain, transform-domain operations simplify subsequent digital processing and improve overall signal robustness.

\subsection{Spatiotemporal Processing}

It has recently been demonstrated that space-time block codes (STBCs), such as the Alamouti code, can be implemented on a traditional RIS by configuring phase shifts over time; this configuration creates transmit diversity effects without requiring additional RF chains \cite{9162097}. In such designs, phase-only modulation of the surface is used to emulate space-time diversity patterns, yielding STBC-like behavior. This principle can be extended to more general STBC structures, where time-varying surface configurations generate diversity patterns analogous to conventional space-time signaling. The same concept naturally applies to transceiver-integrated BD-RIS architectures. Due to their ability to jointly adjust amplitude and phase, BD-RIS provides a richer design space for spatiotemporal processing, enabling more flexible and potentially more faithful wave-domain realizations of space-time coding principles. By synchronizing the time-varying impedance configuration with symbol intervals, transceiver-integrated BD-RIS can implement STBC-like spatiotemporal wave processing directly in the analog domain, reducing reliance on digital space-time encoding with a large number of power-hungry RF chains.

\section{Challenges and Research Agenda}

While transceiver-integrated BD-RIS opens a new frontier in wave-domain signal processing, several challenges need to be addressed before commercialization. Accordingly, unlocking the full potential demands research efforts spanning electromagnetic modeling, circuit design, algorithm development, and experimental validation. This section outlines a research agenda on different directions, including developing physically consistent analytical models, designing low-complexity circuit topologies and configuration algorithms, validating concepts through practical prototypes, and enabling the simultaneous execution of multiple wave-domain functions on a single surface.

\subsection{Physically Consistent Modeling \textcolor{black}{and Hardware Impairments}}

A major research priority is developing physically consistent models that capture \textcolor{black}{both EM interactions and practical hardware impairments.} On the EM side, strong coupling between the active antennas and the BD-RIS elements,  as well as mutual coupling among the array and surface elements must be accurately characterized \cite{11311540, 10530995}. Existing studies often overlook these interactions by assuming simplified transceiver-integrated BD-RIS models, which can lead to misleading performance predictions. Another key aspect is the near field channel between the active antenna and the BD-RIS elements, which is generally approximated using the Rayleigh-Sommerfeld diffraction theory \cite{11301914, raeisi2024efficient, 10515204}. Although this approximation provides valuable insights, it may not accurately reflect actual performance in the presence of reactive near field effects.
\textcolor{black}{On the hardware side, practical implementations are further affected by fabrication imperfections, calibration errors, insertion losses, and finite-resolution impedance tuning. These non-idealities introduce additional model mismatch and performance degradation beyond those predicted by idealized analyses. Therefore, impairment-aware modeling and robust design methodologies are essential for developing practically reliable analog signal processing architectures.}


\subsection{Circuit Complexity and Topology Design}

The scalability of transceiver-integrated BD-RIS depends on its circuit complexity. A high-complex reconfigurable impedance network entails high computational and control overhead \textcolor{black}{in configuring the analog processing network, which may partially reduce the complexity benefits achieved by migrating signal processing from the digital baseband to the analog domain}. Each additional inter-element connection introduces a new impedance path, increasing these overheads, along with the insertion loss and calibration burden. Therefore, future research should focus on designing low-complex circuit topologies \textcolor{black}{that keep the configuration burden manageable while maintaining the overall complexity benefits of analog-domain processing.}

\subsection{Configuration Algorithms and Computational Complexity}

Low-complexity circuit topologies mainly reduce the number of reconfigurable impedance loads; however, they do not determine how difficult it is to obtain an optimal configuration. In fact, algorithmic complexity 
is driven by the underlying signal processing task. Different tasks involve different objective functions and constraints, which naturally lead to different optimization approaches. Moreover, each low-complexity circuit topology introduces its own structural constraints, further shaping the feasible solution space. As a result, designing low-complexity configuration algorithms for BD-RIS cannot follow a one-size-fits-all approach. Instead, future research must jointly consider circuit topology, task objectives, and algorithmic structure to achieve practical, scalable, and efficient BD-RIS configuration.

\subsection{Multi-Function Simultaneity}

A key challenge in transceiver-integrated BD-RIS is enabling \textcolor{black}{joint multi-objective} wave-domain \textcolor{black}{processing, where a single BD-RIS configuration is designed to support multiple functionalities}. A wireless communication system may require \textcolor{black}{concurrent} support for tasks such as communication beamforming, sensing, interference management, and channel conditioning. \textcolor{black}{In this context, BD-RIS does not implement these functions independently; rather, a single configuration is jointly optimized to satisfy multiple objectives within a unified analog transformation.}
Achieving this level of integration requires further research efforts. Specifically, different functions should coexist without detrimental effects, \textcolor{black}{and} optimization frameworks \textcolor{black}{must} jointly configure the BD-RIS response while accounting for these overlapping objectives. Addressing this challenge is essential for transforming BD-RIS from a single-purpose wave-control interface into a fully multi-functional analog processor for next-generation transceivers.

\subsection{Experimental Validation Through Practical Prototypes}

The existing studies on transceiver-integrated BD-RIS primarily rely on theoretical and simulation analysis, while experimental validation is not investigated. \textcolor{black}{Notably, a practical BD-RIS prototype has been reported in \cite{11195964}, demonstrating the feasibility of implementing transmissive BD-RIS. However, their integration into transceiver architectures is still an open challenge. Experimental investigations} are required in order to assess the validity of theoretical models in the presence of hardware impairments, strong coupling effects, imperfect impedance loads, and fabrication shortcomings. Such an experimental investigation can shed light on challenges that are not trivial in simulation studies. Furthermore, experimental investigations are essential to validate physically consistent theoretical models and to assess whether low-complexity circuits and optimization algorithms remain effective in practical environments. Bridging this gap remains a critical step toward practical deployment.

{\color{black}
\section{Conclusion}

This article reviews the shift from conventional digital baseband processing toward analog wave-domain signal processing, with a particular focus on transceiver-integrated BD-RIS architectures. Driven by the scalability limitations of fully digital and hybrid A/D architectures, it discusses key analog processing solutions, namely MiLAC, SIM, and transceiver-integrated BD-RIS, and compares their principles and trade-offs. Among them, transceiver-integrated BD-RIS provides a compelling balance between flexibility, scalability, and performance, making it a strong candidate for next-generation 6G transceivers. Looking ahead, wave-domain signal processing with BD-RISs opens new opportunities for integrated communication, sensing, and localization, while requiring advances in physically consistent modeling, low-complexity circuit design, and scalable configuration algorithms.
}

\ifCLASSOPTIONcaptionsoff
  \newpage
\fi

\bibliographystyle{IEEEtran}
\bibliography{references}

\end{document}